\documentstyle[11pt,epsf]{article}

\begin{document}

\title{
Singular Structure in 4D Simplicial Gravity}

\author{S. Catterall\\
Physics Department, Syracuse University, Syracuse NY 13244\\
R. Renken\\
Physics Department, University of Central Florida, Orlando,
FL 32816\\
J. Kogut\\
Loomis Laboratory, University of Illinois, 1110 W. Green St.
Urbana, IL 61801}

\maketitle

\begin{abstract}
We show that the phase transition previously observed in dynamical triangulation
models of quantum gravity can be understood as being due to the
creation of a {\it singular} link. The transition between singular
and non-singular geometries as the gravitational coupling is varied appears
to be first order. 
\end{abstract}

Dynamical triangulations (DT) furnish a powerful approach to the problem of
defining and studying a non-perturbative theory of quantum gravity. Simply
put, the functional integral over (euclidean) four-geometries is defined as 
some scaling limit of a   
sum over abstract simplicial manifolds. In two dimensions this 
approach has been very successful \cite{revs}. In addition to the
calculation of gravitationally dressed anomalous dimensions, the DT
approach, being discrete, allows for the use of non-perturbative methods
such as computer simulation. In two
dimensions this has yielded new results; the 
measurement of fractal dimensions characterizing the quantum geometry
\cite{fractal},
power law behavior of matter field correlators on geodesic paths \cite{
geodesics}
and insight into the problem of formulating a renormalization group
for quantum gravity \cite{rg}.

The observation of a two phase structure for the four dimensional models
was noticed early on \cite{early4d}. This led to the speculation that
a continuum theory of gravity could be constructed by taking
an appropriate 
scaling limit near 
the phase transition. However, in \cite{first}, evidence was
presented that the transition was {\it discontinuous}. Such a 
scenario would rule out arguably the simplest possibility -- that
a continuum theory could be obtained by tuning the theory to some
critical (gravitational) coupling.

Independent of these issues it was noted in \cite{sing} that the
crumpled phase of this model possessed a singular structure -- that
the sum over triangulations was dominated by those with
a single link common to a very large number of four-simplices. The endpoints
of this link, termed singular vertices, were shared by a number of
four-simplices which diverged linearly with the total triangulation
volume\footnote{Singular vertices 
were first noticed 
in \cite{sing_early}}. 

In this note we have studied how this singular
structure changes as the gravitational coupling is varied. We find evidence that
the transition is driven by the creation of singular vertices as
the coupling is lowered, singular vertices being absent in the
branched polymer (large coupling) phase. We support our
observations by appealing to a mean-field argument due to
Bialas at al \cite{cmf}.

The model we have investigated is defined by the grand
canonical partition
function

\begin{equation}
Z_{GC}=\sum_{T} e^{-\kappa_4 N_4 + \kappa_0 N_0}
\end{equation}

The coupling $\kappa_0$ plays the role of an (inverse) bare gravitational
constant conjugate to a discrete analog of the integrated curvature --
the number of vertices $N_0$.
In practice we use $\kappa_4$ (the bare cosmological constant) to fix 
the volume (number of four-simplices $N_4$) and consider a
canonical partition function 

\begin{equation}
Z_C\left(\kappa_0,V\right)=\sum_{T_V} e^{\kappa_0 N_0}
\end{equation}

The class of triangulations $T_V$ with volume $V$ is further restricted
to be those with the global topology of the four-sphere together with
a local manifold constraint -- that the vicinity of any point should
be homeomorphic to a four-ball.

\begin{figure}[htb]
\centering
\epsfxsize=4.5in \epsfbox{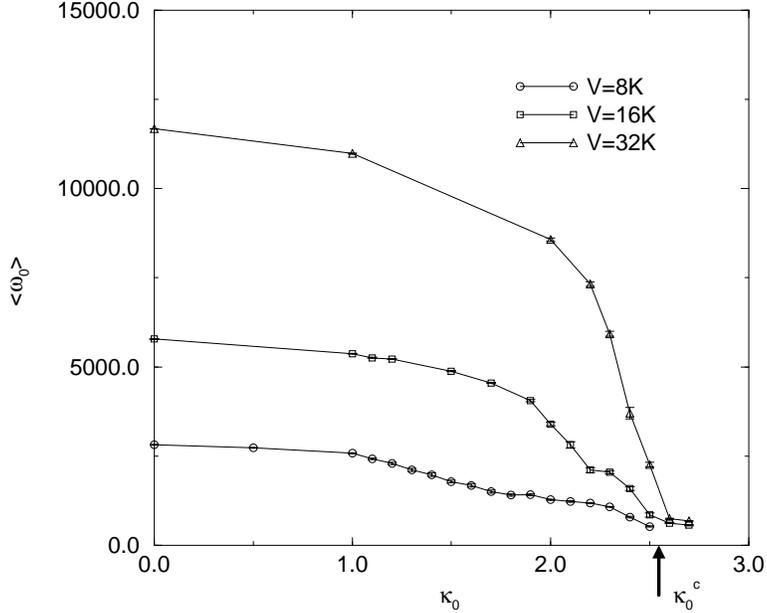}
\caption{Singular vertex volume vs $\kappa_0$} 
\label{fig1}
\end{figure}

We use a Monte Carlo algorithm, described in \cite{code},
to sample the dominant contributions
to $Z_C$. Fig. \ref{fig1} shows a plot of the local volume  of the
most singular vertex $<\omega_0>$ 
(i.e the number of simplices sharing that vertex)
against coupling $\kappa_0$ for lattices with volume $8K$, $16K$ and
$32K$. 

Deep in the crumpled phase it is clear that $<\omega_0>$ scales
linearly with system size as expected. For sufficiently large
coupling $\kappa_0$ it is small, the ratio $<\omega_0>/V$ approaching zero
as $V\to\infty$. The  behavior of $\omega_0$
for intermediate coupling is somewhat
complex -- on the $8K$ lattice it falls rather slowly for $\kappa_0<1$,
kinks for $\kappa_0\sim 1$, falls more rapidly until
$\kappa_0\sim 1.8$ and then plateaus until $\kappa_0\sim 2$.
We will
call the regime from $\kappa_0\sim 1$ to $\kappa_0\sim 2$
the {\it mixed region}. For couplings larger than
this $<\omega_0>$ rapidly approaches its asymptotic value.
A similar picture is seen at $16K$ -- the volume $\omega_0$ falls
slowly up to $\kappa_0\sim 2$ after which it falls quickly to a
small plateau for $\kappa_0\sim 2.2$. For couplings larger than this
the value of $\omega_0$ quickly approaches its limiting value. Thus the
behavior is similar to the $8K$ data except that
the mixed region is now much smaller.
The arrow on the plot indicates the approximate position of the
usual phase transition as revealed by measurements of the
vertex susceptibility $\frac{1}{V}\left(\left<N_0^2\right>-
\left<N_0\right>^2\right)$ for large volumes. It appears that
the rightmost boundary of the mixed region coincides with the
phase transition. At $32K$ the mixed region has narrowed to the point where
it can no longer be resolved and the data is best
interpolated with a curve that undergoes a rapid variation in its gradient
close to the phase transition.

\begin{figure}[htb]
\centering
\epsfxsize=4.5in \epsfbox{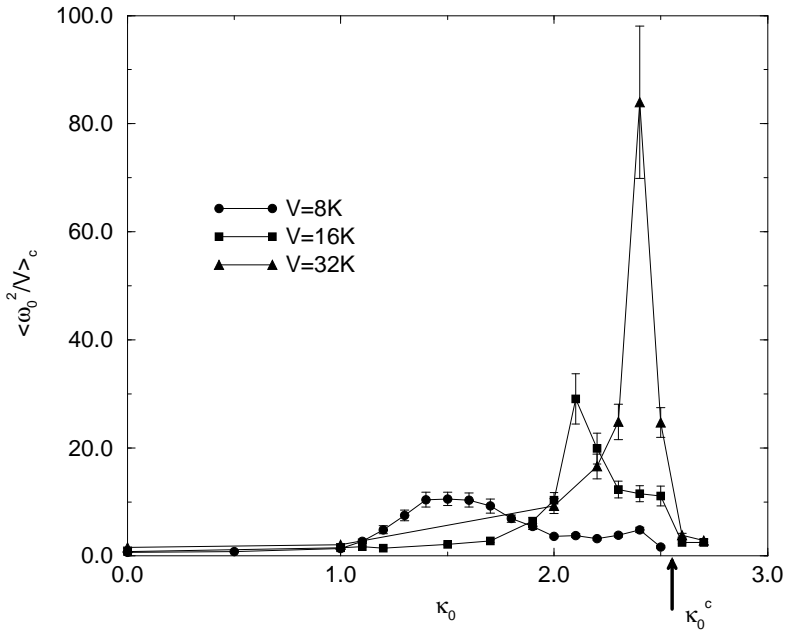}
\caption{Fluctuations in singular vertex volume $\chi_0$} 
\label{fig2}
\end{figure}

Further insight can be gained by looking at the fluctuations in this
local volume 
$\chi_0=\frac{1}{V}\left(\left<\omega_0^2\right>-\left<\omega_0\right>^2
\right)$. In Fig. \ref{fig2} it is clear that the data for small
volume lies on a curve with two peaks -- corresponding to the boundaries
of the mixed region. These peaks appear to merge on the
phase transition point as the volume is increased. For $32K$ these two
peaks have already coalesced into one.

\begin{figure}[htb]
\centering
\epsfxsize=4.5in \epsfbox{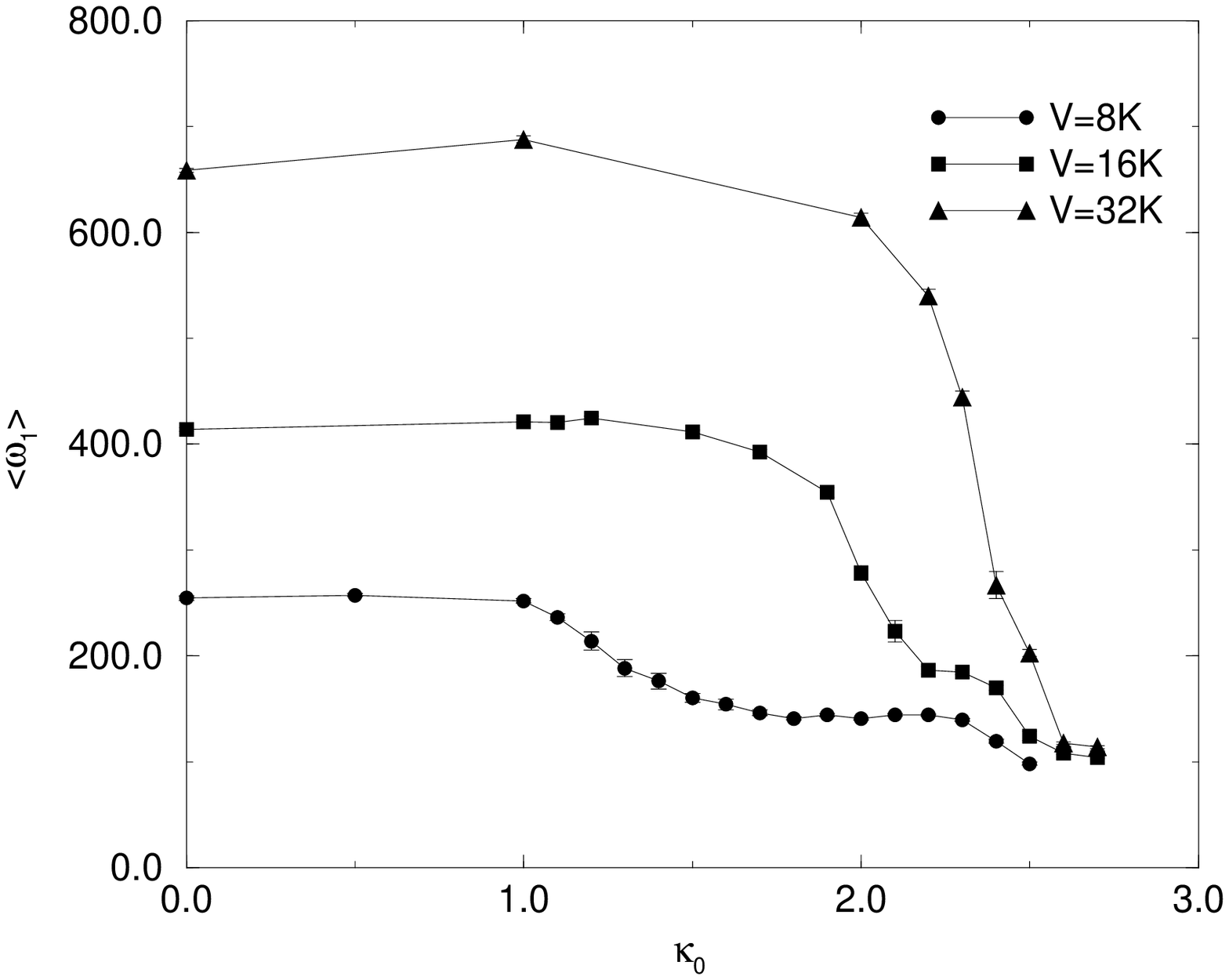}
\caption{Singular link volume vs $\kappa_0$} 
\label{fig3}
\end{figure}

\begin{figure}[htb]
\centering
\epsfxsize=4.5in \epsfbox{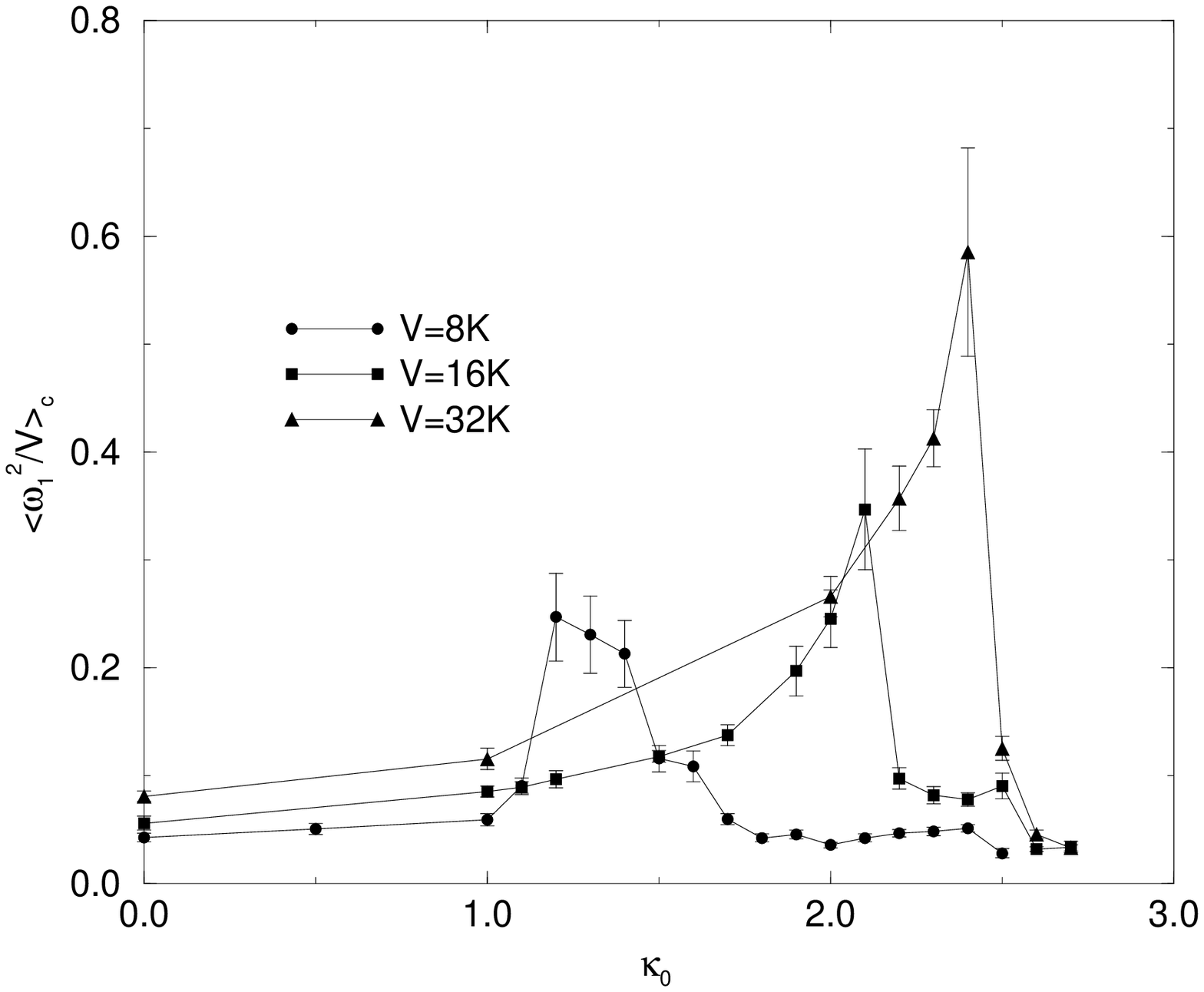}
\caption{Fluctuations in singular link volume $\chi_1$} 
\label{fig4}
\end{figure}

The explanation for these finite size effects can be understood by
plotting the maximal link local volume $<\omega_1>$ (fig. \ref{fig3}). The
left edge of the mixed region $\kappa_0^1(V)$
matches an abrupt change in $<\omega_1>$ as
$\kappa_0$ is increased - indeed by looking in detail at the
triangulations we find for couplings $\kappa_0$ in the mixed region the
link between the singular vertices has broken -- there is no identifiable
singular link - merely a gas of a variable number of remnant
singular vertices (one, two or more vertices with
large local volumes). The rightmost boundary of the mixed region
$\kappa_0^2(V)$ then 
corresponds to the final disappearance of the
remaining singular vertices. This conclusion is strengthened by looking at the
conjugate susceptibility $\chi_1=\frac{1}{V}\left(\left<\omega_1^2
\right>-\left<\omega_1\right>^2\right)$ (fig. \ref{fig4}). The peak
associated with the breakup of the singular link appears to
merge with the peak associated with the disappearance of singular
vertices as the volume gets larger.  

To summarize, our study of the
singular structure indicates that at finite volume there are
two pseudo-critical couplings $\kappa_0^1$ and $\kappa_0^2$ 
associated to the creation of singular links and vertices 
respectively. These appear to lock into a single critical 
point $\kappa_0^c$  as $V\to\infty$\footnote{Similar results were obtained
independently by B. Petersson and J. Tabaczek 
\cite{taba}}.
Furthermore, 
the gradient of the $<\omega_0>$ curves in
the transition region increases with volume $V$. This is
consistent with a finite discontinuity of $<\omega_0>$ in the
infinite volume limit. Indeed if we interpret $<\omega_0>/V$ as an
order parameter then we can read off the order of the transition
from the scaling of its corresponding susceptibility $\chi_0$. The
numerical data is certainly consistent with a linear scaling of
the peak in $\chi_0$ with volume, characteristic of a
first order phase transition.

If we run simulations close to $\kappa_0^c$ for large volumes we
see signs of meta-stability -- the system tunnels back and
forth between two distinct states -- one is branched polymer-like
and contains no singular vertex and the other contains
one or more remnant singular vertices. This type of behavior
has been seen before and is a strong indicator of a discontinuous
phase transition.

\begin{figure}[htb]
\centering
\epsfxsize=4.5in \epsfbox{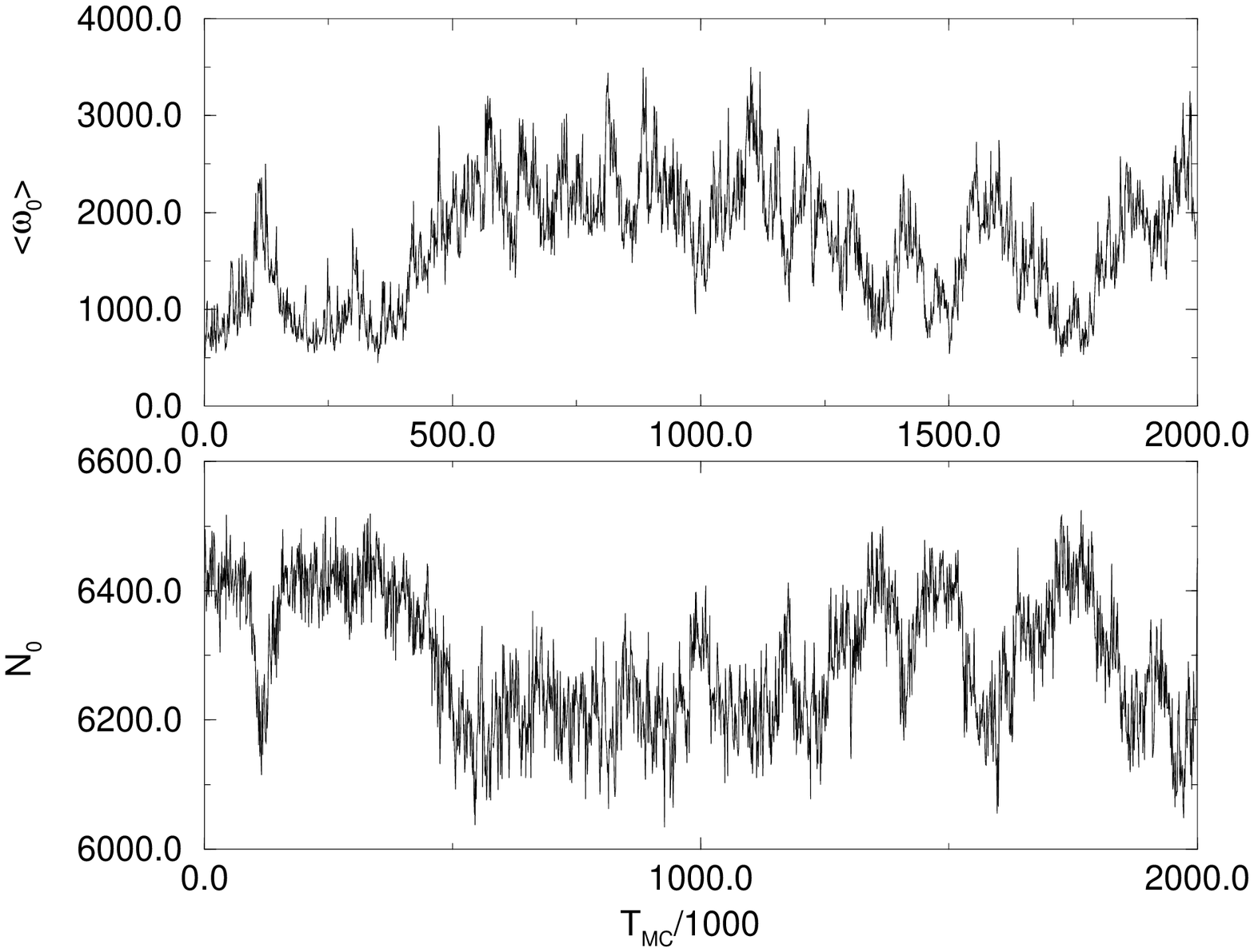}
\caption{Correlation of vertex number $N_0$ with singular vertex volume} 
\label{fig5}
\end{figure}

The correlation between the action $N_0$ and singular vertex volume
$\omega_0$ can be seen easily in fig. \ref{fig5} where time
series of the two quantities
are plotted. There is clearly a perfect (anti)correlation - 
fluctuations to larger vertex numbers are accompanied by
precisely matching decreases in the singular vertex volume. This
is yet another indicator that the physics of the phase transition
corresponds exactly to the fluctuations in the singular
structure.  

It is possible to gain further insight into the nature of this
transition by making a couple of assumptions. Consider the microcanonical
partition function $Z_{MC}\left(V,N_0\right)$
gotten by fixing the number of vertices. Following
\cite{sing} let us define a local entropy $s_i$ associated with a
vertex $i$. This is just the number of ways of
gluing together the simplices in its local
volume element -- hence it is a function
of that local
volume $s_i=s_i(\omega_0(i))$. We will first assume 
that 
$Z_{MC}$ can be approximated by enumerating all possible ways in
which these local entropies can be assigned\footnote{ In the light of
the results derived in [11], we should probably regard this as
an {\it effective} theory valid for understanding some of
the properties of the phase transition -- it seems that the leading
behavior of the partition function can be derived from a  
partitioning of simplices amongst the (non-singular) $(d-2)$-simplices --
the effect of the singular structure being sub-leading.}.
This is a drastic
assumption but we are encouraged by the results of
our simulations which support the notion that the transition is
driven by fluctuations in the singular vertex 
structure. Secondly we will assume that for sufficiently small
vertex density $N_0/V$ a mean field approximation can be used in which the
local entropies are treated as independent. Thus
we are led to an ansatz for the partition function
of the form

\begin{equation}
Z_{MC}\left(V,N_0\right)=\sum_{\{\omega_0(i)\}}\prod_i^{N_0} s_i\left(
\omega_i(i)\right)\delta\left(\sum_i
\omega_0(i)-cV\right)
\end{equation}

The final delta function reflects the fact that at least
one global constraint restricts the local volumes $\omega_0(i)$ 
-- that the sum of these volumes is just proportional to the
volume $V$. ($c$
the constant of proportionality is equal to five). As argued in
\cite{sing} for $s_i\left(\omega_0(i)\right)$ we should take the
number of triangulations of the three sphere bounding 
the vertex $i$. 
The precise functional form of $s_i$ is unknown but
previous numerical simulations \cite{3d} and
an analytic calculation of the exponential bound \cite{mauro}
lead to the form
 
\begin{equation}
s_i\left(n\right)\sim \exp{(-an^\sigma)}e^{bn}
\end{equation}
\noindent

The constant $\sigma\sim 0.6-7$. 
Using the results of \cite{cmf} one can write the partition function of
this model in the thermodynamic limit as

\begin{equation}
Z_{MC}=\frac{1}{2\pi}\exp{\left(\left(-\log{\lambda}+b\right)cV +
                          \ln{F\left(\lambda\right)}N_0\right) }
\label{cmf_eq}
\end{equation}

\noindent
where the parameter $\lambda$ is the solution of the equation

\begin{equation}
\rho=\frac{\lambda F^\prime\left(\lambda\right)}{F\left(\lambda\right)}
\end{equation}

\indent
and

\begin{equation}
F\left(\lambda\right)=\sum_{\omega}s^\prime\left(\omega\right)\lambda^\omega
\end{equation}

The prime indicates that the leading exponential piece in $s_i$ has
been trivially factored out (it can be seen in the leading term in
eqn.~\ref{cmf_eq}) and so $F$ depends only on the sub-exponential
pieces in $s_i$.

The model exhibits a two phase
structure; for densities $\rho=cV/N_0$ with
$\rho> \rho_c=\frac{F^\prime\left(1\right)}{F\left(1\right)}$
the parameter $\lambda=1$ and the system is in a collapsed phase where vertices
of small order behave independently and the global constraint is satisfied by a
small number of vertices which are shared by a number of simplexes
on the order of the volume. This regime is
identified with the crumpled phase of the dynamical triangulation model.
Conversely for $\rho< \rho_c$ the parameter $\lambda$ varies between
zero and one and the free energy
varies continuously with $\rho$. The distribution of vertex orders then behaves
as

\begin{equation}
p\left(n\right)\sim \omega\left(n\right)e^{-n\log{\lambda}}
\end{equation}

This phase is then identified with the branched polymer phase in the dynamical
triangulation model. As $\rho$ (and hence $N_0$) is varied (by varying the
coupling $\kappa_0$) the system moves between these two phases. This
phase structure is seen for any value of the parameter
$\sigma< 1$. Indeed any sub-exponential or
power law behavior will suffice. 

Our simulations
are run in the canonical ensemble where a sum over vertices is taken

\begin{equation}
Z_{C}=\sum_{N_0} Z_{MC}\exp{\kappa_0 N_0}
\end{equation}

If we use the mean-field approximation for $Z_{MC}$ this integration
can be done \cite{cmf_first}. The model again exhibits a two phase
structure with a {\it first order} phase transition separating the
crumpled phase from the branched polymer phase. This then is in
agreement with the results of our simulations.
 
In our mean field arguments we have only
considered the vertex sector of the model -- clearly the
links can be considered as possessing a local entropy too. In this case this
entropy would be equal to the number of triangulations of the
two-sphere dual to the link. Similar arguments would then lead to
the prediction of a two phase structure for the links - a collapsed
phase with singular links and a fluid phase in which the link
distribution behaves exponentially with link local volume. The 
coupling $\kappa_0$ at which this transition occurred would
seemingly be independent of the coupling at which singular vertices appear.
Although our simulations are consistent with this scenario at small
volume it appears that these two transition points merge in the
infinite volume limit. Of course, the vertex and link sectors are
not truly independent and it is presumably the presence of these 
residual constraints that is
responsible for the coincidence of the two transitions. 

In summary, we have reported numerical results which strongly support
the idea that the phase transition observed in 4d simplicial gravity
is associated with the creation of singular geometries. These
singular structures, composed of
a single singular link between two singular 
vertices, dominate at strong coupling. As the coupling
$\frac{1}{\kappa_0}$ is decreased they eventually disappear. This
process appears to be discontinuous and is the origin of the
first order nature of the transition. Simple mean field arguments 
support these conclusions.

This work was supported in part by NSF Grant PHY-9503371 and
DOE Grant DE-FG02-85ER40237. We acknowledge useful conversations
with B. Petersson.

\vfill


\vfill

\newpage

\begin{thebibliography}{99}
\bibitem{revs}
 F. David, ``Simplicial Quantum Gravity and Random
 Lattices,'' (hep-th/9303127), Lectures given at Les
 Houches Summer School on Gravitation and Quantizations, 
 Session LVII, Les Houches, France, 1992, \\
 J. Ambj\o rn, ``Quantization of Geometry.'' (hep-th/9411179),
 Lectures given at Les Houches Summer School on Fluctuating 
 Geometries in Statistical Mechanics and Field Theory, 
 Session LXII, Les Houches, France, 1994,  \\
 P. Ginsparg and G. Moore, ``Lectures on 2D Gravity and 2D
 String Theory,'' (hep-th/9304011), Lectures given 
 at TASI Summer School, Boulder, CO, 1992, \\
 P. Di Francesco, P. Ginsparg and J. Zinn-Justin,
 ``2-d Gravity and Random Matrices,''  (hep-th/9306153),
  Phys. Rep. 254 (1995) 1. 
\bibitem{fractal}S. Catterall, G. Thorleifsson, M. Bowick and V. John,
Phys. Lett. B354 (1995) 58.\\
J. Ambj\o rn, J. Jurkiewicz and Y. Watabiki, Nucl. Phys.
B454 (1995) 313.
\bibitem{geodesics} J. Ambjorn and K. Anagnostopoulos, Phys. Lett. B388
(1996) 713.
\bibitem{rg}  G.
Thorleifsson and S. Catterall, Nucl. Phys. B461 (1996) 350.\\
R. Renken, Phys. Rev. D50 (1994) 5130.
\bibitem{early4d} M. Agishtein and A. Migdal, Nucl. Phys. B385 (1992) 395.\\
J. Ambj\o rn and J. Jurkiewicz, Phys. Lett. B278 (1992) 42. \\
B. Brugmann and E. Marinari, Phys. Rev. Lett 70 (1993) 1908.\\
S. Catterall, J. Kogut and R. Renken, Phys. Lett. B328 (1994) 277.\\
B. de Bakker and J. Smit, Nucl. Phys. B437 (1995) 239. 
\bibitem{first} P. Bialas, Z. Burda, A. Krzywicki and B. Petersson 
 Nucl. Phys. B472 (1996)
293.\\
B. de Bakker, Phys. Lett. B389 (1996) 238.
\bibitem{sing}
S. Catterall, G. Thorleifsson, J. Kogut and R. Renken, Nucl. Phys. B468 (1996).
\bibitem{sing_early} T. Hotta, T. Izubuchi and J. Nishimura, Nucl.Phys.Proc.Suppl. 47 (1996) 609-612.
\bibitem{cmf} P. Bialas, D. Johnston and Z. Burda, Nucl. Phys. B493 (1997) 505.\\
P. Bialas, Z Burda, B. Petersson and J. Tabaczek, Nucl. Phys. B495 (1997) 463.
\bibitem{code}
S. Catterall, Computer Physics
Communications 87 (1995) 409.
\bibitem{taba} J, Tabaczek: Diplomarbeit Bielefeld Juni 1997.
\bibitem{3d}  S. Catterall, R. Renken and J. Kogut, Phys. Lett. B342 (1995)
53.\\
J. Ambjorn and S. Varsted, Phys. Lett. B226 (1991) 285.
\bibitem{mauro} ``The Geometry of Dynamical Triangulations", J. Ambjorn, M
Carfora and A. Marzuoli., Springer Verlag to be published.
\bibitem{cmf_first} P. Bialas and Z. Burda, hep-lat/9707028.

\end{thebibliography}
\end{document}